\documentstyle[12pt,epsf,psfig]{article}
\begin{document}
\setlength{\headheight}{0in}
\setlength{\headsep}{0in}
\setlength{\topskip}{1ex}
\setlength{\textheight}{8.5in}
\setlength{\topmargin}{0.5cm}
\setlength{\baselineskip}{0.24in}
\catcode`@=11
\long\def\@caption#1[#2]#3{\par\addcontentsline{\csname
  ext@#1\endcsname}{#1}{\protect\numberline{\csname
  the#1\endcsname}{\ignorespaces #2}}\begingroup
    \small
    \@parboxrestore
    \@makecaption{\csname fnum@#1\endcsname}{\ignorespaces #3}\par
  \endgroup}
\catcode`@=12
\def\slashchar#1{\setbox0=\hbox{$#1$}           
   \dimen0=\wd0                                 
   \setbox1=\hbox{/} \dimen1=\wd1               
   \ifdim\dimen0>\dimen1                        
      \rlap{\hbox to \dimen0{\hfil/\hfil}}      
      #1                                        
   \else                                        
      \rlap{\hbox to \dimen1{\hfil$#1$\hfil}}   
      /                                         
   \fi}                                         %
\newcommand{\newc}{\newcommand}
\def\be{\begin{equation}}
\def\ee{\end{equation}}
\def\bea{\begin{eqnarray}}
\def\eea{\end{eqnarray}}
\def\simlt{\stackrel{<}{{}_\sim}}
\def\simgt{\stackrel{>}{{}_\sim}}
\begin{titlepage}
\begin{flushright}
{\setlength{\baselineskip}{0.18in}
{\normalsize
IC/99/62\\
hep-ph/9905xxx\\
May 1999\\
}}
\end{flushright}
\vskip 2cm
\begin{center}

{\Large\bf 
Additional phases induced by the supersymmetric CP phases
}

\vskip 1cm

{\large
D. A. Demir \\}

\vskip 0.5cm
{\setlength{\baselineskip}{0.18in}
{\normalsize\it The Abdus Salam International Center for Theoretical Physics, I-34100,
Trieste, Italy \\}}

\end{center}
\vskip .5cm
\begin{abstract}
The explicit CP violation in the MSSM radiatively induces a finite unremovable 
alignment between the Higgs doublets. This additinal phase can be as large as the
original CP phases in certain portions of the MSSM parameter space. Considering 
the specific case of the charginos, this additional phase is shown to induce a conceivable amount of CP 
violation near the would--be CP conserving points. Moreover, the CP violation in the absence of 
this phase is smaller than the one in the presence of it, and the former can never compete 
with the latter, however large  $\tan\beta$ is.

\end{abstract}
\end{titlepage}
\setcounter{footnote}{0}
\setcounter{page}{1}
\setcounter{section}{0}
\setcounter{subsection}{0}
\setcounter{subsubsection}{0}
\section{Introduction}
The Lagrangean of the  Minimal Supersymmetric Standard Model (MSSM) consists of 
various mass parameters which are not necessarily real. Indeed, as  
was studied in \cite{phase1,phase2}, after all possible rephasings of the fields
there remain two physical phases which can be chosen to be those of 
the $\mu$ parameter ($\varphi_{\mu}\equiv \mbox{Arg}[\mu]$), and Higgs--sfermion--sfermion
trilinear couplings ($\varphi_{A_{f}}\equiv \mbox{Arg}[A_{f}]$). Of course, in addition 
to these, the QCD vacuum angle $\theta_{QCD}$ and the phase in the CKM 
matrix $\delta_{CKM}$ are still present as in the Standard Model (SM).

When analyzing their effects on the low--energy processes one may regard these phases as given 
parameters in the MSSM Lagrangian \cite{phase1,phase2} altough it is possible to realize 
them, for example, by embedding the SM on  D--branes \cite{stringy} or more general
dynamical supersymmetry breaking scenarios \cite{phase2,sugra}. However, independent 
of specific high energy realizations, the low--energy Lagrangian happens to depend on 
several soft phases whose physical combinations can always be chosen to be those of the $\mu$-- 
and $A$--parameters. These phases have found interesting applications in various areas of the particle
phenomenology: EDM's of particles \cite{edm}, LSP searches \cite{falk},
$K$-- and $B$-- physics \cite{phase1,others}, electroweak baryogenesis \cite{baryo}, weak and 
electromagnetic dipole formfactors \cite{dipole}, and Higgs phenomenology \cite{ben,wagner}.

In the presence of these phases the CP--invariance of the MSSM Lagrangian is explicitly
broken, and naturally, various interaction vertices get modified. Moreover, through the
radiative corrections, the squared--mass matrix of the Higgs scalars assumes CP--violating
elements thereby the mass--eigenstate Higgs scalars turn out to have no definite CP quantum
numbers \cite{ben,wagner}. Furthermore, due to CP--violation, these phases induce 
a finite alignment between the Higgs doublets \cite{ben,wagner} which would be absent 
if there were no CP--violation, explicit or spontaneous \cite{spon}. This additional phase can have 
non--trivial effects on the interactions of charginos, neutralinos and sfermions.
In what follows, its size as well as its effects on the mixing matrices of certain particles
will be analyzed as a function of the supersymmetric parameter space. 

In the analysis below, $\varphi_{\mu}$ and $\varphi_{A_{f}}$ will be taken free, that is,
it is assumed that the electric dipole moment contraints \cite{phase1,edm} could be sidestepped
by following the appropriate parameter spaces discussed in \cite{side}. In this work, sfermions
of the first two generations will be assumed degenerate and very heavy (see the last three
references in \cite{side}) compared to the third generation sfermions. With such heavy sfermions
belonging to the first two generations one can suppress the electric dipole moments of the particles
in agreement with the experimental bounds.  

The radiatively induced unremovable alignment between the Higgs doublets can have nontrivial effects on
the chargino and neutralinos sectors. For both clarity and simplicity this work deals mainly with the 
chargino sector. Accordingly, analyses below will be based on the typical combinations of the chargino mixing matrices 
appearing in the relevant processes listed above. Section II discusses (1) the induction and size of 
the additional CP phase, and (2) its effects on the chargino sector. Section III concludes the work.

\section{Radiatively Induced CP Phases}
The MSSM Higgs sector contains two opposite--hypercharge Higgs doublets $H_{1}$, $H_{2}$ in 
terms of which the tree--level potential reads as follows
\bea
V_{0}(H_{1}, H_{2})&=&m_{1}^{2}|H_{1}|^2+m_{2}^{2}|H_{2}|^2+
  (m_{3}^{2}H_{1}\cdot H_{2}+H.c.)\nonumber\\
&+&\frac{\lambda_{1}}{2}|H_{1}|^{4}+\frac{\lambda_{2}}{2}|H_{2}|^{4}+
\lambda_{12} |H_{1}|^2 |H_{2}|^2+\tilde{\lambda}_{12}|H_{1}\cdot H_{2}|^{2}   
\eea
where the parameters are defined by 
\bea
m_{1}^{2}&=&m_{\tilde{H}_{1}}^{2}+|\mu|^{2}\; , \; m_{2}^{2}=m_{\tilde{H}_{2}}^{2}+|\mu|^{2}\; , \;
\lambda_{1}=\lambda_{2}=(g_{2}^{2}+g_{1}^{2})/4\nonumber\\
\lambda_{12}&=&(g_{2}^{2}-g_{1}^{2})/4\; , \; \tilde{\lambda}_{12}=-g_{2}^{2}/2 \;.
\eea
Here $m_{\tilde{H}_{1,2}}^{2}$ and $m_{3}^{2}$ are the soft supersymmetry breaking masses determining the
Higgs bilinears. Since $\mu$ enters through $|\mu|^{2}$ only, and phase of $m_{3}^{2}$ is rotated away 
already, the potential is spanned by the real parameters only. Therefore, Higgs fields develop 
only real vacuum expectation values. Indeed, even if one introduces a certain alignment $\theta$ 
between  the Higgs doublets $H_{1}$ and $H_{2}$, tadpole equations nullify it \cite{sher}. Namely, in the minimum, 
the potential is to have vanishing gradients in all directions, in particular,  
$\partial V_{0}/\partial \varphi_{1,2} = -m_{3}^{2} \sin \theta \equiv 0$. Here and in what follows 
$H^{0}_{1,2}$  show the neutral components of the Higgs fields, and they are linearly decomposed as 
follows: $H_{1}^{0}=(v_{1}+\phi_{1}+i \varphi_{1})/\sqrt{2}$, $H_{2}^{0}=(v_{2}+\phi_{2}+i \varphi_{2})/\sqrt{2}$ with
$v_{2}/v_{1}\equiv \tan\beta$ and $M_{W}^{2}=g^{2}(v_{1}^{2}+v_{2}^{2})/4$.
   
It is now a well--established fact that the radiative corrections to the MSSM Higgs sector are important 
in that the mass of the lightest Higgs gets a large correction : $\delta m_{h}^{2}\sim M_{Z}^{2}$.
This sizable radiative correction to the Higgs mass has been computed in the framework of
on-shell renormalization \cite{on-shell}, effective potential approach \cite{effpot}, and
RGE--improved one-- and two--loop Higgs potential \cite{RGE} with differing precisions and 
approximation methods. Here, for convenience, effective potential approximation will be followed
as in \cite{ben} where the entire one--loop effects of the MSSM particle spectrum were 
approximated by the dominant top quark and top squark loops. This approximation is good for
picking up phase--dependent dominant terms as long as very large values of $\tan\beta$ are
avoided: $\tan\beta\simlt 40$.  If larger $\tan\beta$ values are considered bottom--sbottom 
and tau--stau loops start dominating, and the approximation adopted here fails. In what follows,
analytic results of \cite{ben} will be frequently referred to avoid unnecessary repetations.

When the one--loop corrections are added to the tree--level potential (1), 
it is seen that $\partial V/\partial \varphi_{1,2}$ no longer vanish; hence, one has to 
redefine the Higgs doublets with a relative phase between them, say,
$H_{1}\rightarrow H_{1}, H_{2}\rightarrow e^{i \theta} H_{2}$ with which now 
$\partial V/\partial \varphi_{1,2}\equiv 0$ gives \cite{ben}
\bea
\sin \theta =-\beta_{h_{t}} \frac{|\mu| |A_{t}|}{M_{A}^{2}\sin 2\beta} \sin \gamma
f(m_{\tilde{t}_{1}}^{2}, m_{\tilde{t}_{2}}^{2})
\eea
where $\beta_{h_{t}}=h_{t}^{2}/16 \pi^{2}$, $\gamma = \varphi_{\mu}+\varphi_{A_{t}}$, and 
$M_{A}^{2}=-m_{3}^{2}/(\sin\beta \cos\beta)$ corresponds to the tree--level pseudoscalar mass
in the CP--conserving theory. Finally, the loop function $f$ is defined by 
\bea
f(x,y)=-2+\log{\frac{xy}{Q^{4}}}+\frac{ y + x } { y - x} \log{\frac{y}{x}} 
\eea
which has an explicit dependence on the renormalization scale $Q$. The stop masses $m_{\tilde{t}_{1,2}}$
in (3) are given by 
\bea 
m_{\tilde{t}_{1(2)}}^{2}=\frac{1}{2}\left ( M_{\tilde{L}}^{2}+M_{\tilde{R}}^{2}+ 2
m_{t}^{2} -(+) \Delta_{\tilde{t}}^{2}(\gamma)\right)
\eea
where $\gamma=\gamma_{\mu}+\gamma_{A_{t}}$, and
\bea
\Delta_{\tilde{t}}^{2}(\gamma)=\sqrt{(M_{\tilde{L}}^{2}-M_{\tilde{R}}^{2})^{2}+ 4 m_{t}^{2} (
|A_{t}|^{2}+
|\mu|^{2} \cot^{2} \beta-2 |\mu| |A_{t}| \cot \beta \cos \gamma)}\; .
\eea
Before going into details it is convenient to dicuss the role of the parameter $M_{A}$ in eq. (3).
As mentioned just after (3) it is the tree level pseudoscalar mass in the CP--conserving theory, that
is, it does not include the loop corrections. Indeed, as derived in \cite{ben}, the squared--mass matrix
of the Higgs scalars depends on the combination $\tilde{M}_{A}^{2}=M_{A}^{2} (\sin(\gamma-\theta)/\sin\gamma)$
instead of $M_{A}^{2}$. However, even $\tilde{M}_{A}$ is itself away from representing the mass of any of the scalars;
there are further radiative corrections determining the Higgs masses and mixings. Altough $\tilde{M}_{A}$ is an
appropriate fundamental variable for the Higgs sector, in studying the relative phase $\theta$ it is convenient to take
$M_{A}$ fundamental as this causes no physical difficulty as long as one deals with variables having explicit dependence on
the renormalization scale $Q$. 
  
The second important thing to be noted is the explicit $Q$ dependence of $\theta$. Usually the explicit 
$Q$ dependence is assumed to cancel with the implicit $Q$ dependencies of $\tan\beta$ and $m_{3}^{2}$ \cite{effpot}.
Besides, in analyzing the Higgs sector this explicit $Q$ dependence is embedded into the quantity $\tilde{M}_{A}$
mentioned above. However, in general, all tree--level quantities are to be interpreted as evaluated at the scale $Q$.
Indeed, the logarithms produced by the renormalization group evolution of the tree level quantities from GUT to the weak
scale have been shown to slow down the $Q$ dependence coming from the effective potential \cite{stable}.
In the following $Q$ will be taken at the weak scale, that is,  $Q\sim m_{t}$. When the supersymmetric mass parameters
are around $m_{t}$, as will be the case in the following numerical studies, the $Q$--dependence of $\theta$ will be weak.

For further progress, it is convenient to clarify the dependence of $\sin\theta$ on various
MSSM parameters on the right--hand side of (3):
\begin{enumerate}
\item $\sin\theta$ vanishes identically once a CP--conserving point is approached, that is,
$\gamma\rightarrow 0, \pi$. Therefore, $\theta$ is solely induced by the CP violating phases,
and vanishes whenever the CP--invariance is restored. 
\item $\sin\theta$ varies as $1/M_{A}^{2}$, and thus, it is diminished in the 
decoupling regime \cite{decoupling}.
\item Putting aside the effects of the loop function, $\sin\theta\propto \tan\beta /\sin^{2}\beta $
so that it grows with increasing $\tan\beta$.
\item For large $\tan\beta$, $\gamma$ dependence of $\Delta_{\tilde{t}}^{2}(\gamma)$ weakens 
and $\sin\theta$ becomes proportional to $\sin\gamma$. 
\item  $\sin\theta\propto |\mu||A_{t}|$, and the latter could, in principle, be large. Here 
the only constraint on $|\mu|$ and $|A_{t}|$ arises from the requirement of keeping
electric and color symmetries unbroken \cite{ccb}. This requires the light stop squared--mass be positive, 
or best, be above the existing experimental lower bounds. 
\item As the detailed discussions in \cite{ben,wagner} show masses as well as the couplings of the MSSM
particles now depend on the phase $\gamma$, as a result, various constraints on the parameter space, derived
for $\gamma=0$, do not necessarily hold for finite $\gamma$. Thus, one can, in fact, relax certain mass bounds
especially in the Higgs sector due to strong $\gamma$ dependence of their masses and couplings, and also 
their indefinite CP quantum numbers. 
\end{enumerate}

In the light of above--listed properties, one concludes that $\sin\theta$ could be 
of the same order as $\sin\gamma$ if 
\bea
\frac{|\mu||A_{t}|}{M_{A}^{2} \sin^{2}\beta \cos\beta} f(m_{\tilde{t}_{1}}^{2}, m_{\tilde{t}_{2}}^{2}) \sim \frac{16}{3} \pi^{2} \, .
\eea
To see if this rough condition is satisfied it is convenient to have a numerical study of it. 
However, MSSM parameter space is too wide to cover fully so it is necessary to choose certain values for the 
parameters meeting, at least, some of the related phenomenological constraints. The induced phase $\theta$ will eventually 
find applications in various processes mentioned in the Introduction, in particular, in electroweak baryogenesis \cite{baryo} and $K$ and
$B$--meson studies \cite{others} both of which  require the soft parameters as well as the $\mu$ parameter be around the weak scale. To
illustrate a conservative case  one can take, for example,  $M_{\tilde{L}}=M_{\tilde{R}}= 2\cdot M_{Z}$ with $m_{\tilde{t}_{1}}\geq
M_{Z}$. Such low values of the stop soft masses together with the lower bound on the light stop mass constrains the stop left--right 
mixings considerably. Hence, $|A_{t}|$ and $|\mu|$ parameters are rather restricted. Despite these constraints, 
if one can find a solution giving $|\theta|\sim {\cal{O}}(1)$, then it is guaranteed that for higher values of the stop left--right
mixings a satisfactory solution will exist. The natural renormalization scale is the weak scale; hence the choice $Q^{2}=
m_{t}^{2}$. $Q$ will always be kept at this value irrespective of the variations in other parameters. 
Apart from these assignments of the parameters, to guarantee that $\sin\theta$ is really large in
some region of the parameter space, it is necessary to compare it with maximal value of $\sin\gamma$, that is, $\gamma=\pi/2$.
In the analysis of $\theta$ below, the remaining parameters $\tan\beta$, $M_{A}$, $|A_{t}|$ and $|\mu|$ will be varied freely
except for the fact that all mass parameters, including the light stop mass, will be required to lie above $M_{Z}$.

Depicted in Fig. 1 is the variation of $\sin\theta$ in $|A_{t}|/M_{Z}$--$|\mu|/M_{Z}$ plane for $\tan\beta=2$ and $M_{A}=M_{Z}$. 
In this as well as other figures $\sin\theta=0$ plateau $is$ $not$ the actual value of $\sin\theta$ instead it shows the $excluded$
$region$ in the $|A_{t}|/M_{Z}$--$|\mu|/M_{Z}$ plane for which either $m_{\tilde{t}_{1}}< M_{Z}$ or $|\sin\theta|>1$. As the figure
suggests, $\sin\theta$ starts with $\sim -0.05$ at $|A_{t}|=|\mu|=M_{Z}$, and falls to $\sim - 0.6$ until $|\mu|\sim 6 M_{Z}$ and
$|A_{t}|\sim 3.2 M_{Z}$. Here $\sin\theta$ cannot reach $-1$ since $m_{\tilde{t}_{1}}$ falls below $M_{Z}$ at the boundary shown.
Thus, for $\tan\beta=2$ and $M_{A}=M_{Z}$, $|\sin\theta|$ value as large as $\sim 60\% |\sin\gamma|$ are reachable. 

Fig. 2 illustrates the same quantity in Fig. 1 for $\tan\beta=30$. Contrary to Fig. 1, here $|\sin\theta|$ exceeds unity well 
before $m_{\tilde{t}_{1}}$ falls below $M_{Z}$ hence the different shapes for the onset of the forbidden regions in two figures. 
It is clear that this time $|\sin\theta|$ reaches unity for $|A_{t}|\sim |\mu|\sim 140\, \mbox{GeV}$. As one recalls from Item (3) in the 
previous page, it is the strong $\tan\beta$ dependence of $|\sin\theta|$ that causes this difference between Figs. 1 and 2. 

Fig. 3 has the same conventions in Fig. 1 but uses $M_{A}=2\cdot M_{Z}$. As already mentioned in Item (2) in the previous page,
$|\sin\theta|$ is rapidly diminished with increasing $M_{A}$. As the figure shows doubling of $M_{A}$ reduces $|\sin\theta|$ from 
the maximal value of $\sim 0.6$ to $0.15$ as is obvious from its $M_{A}$ dependence. 

Finally, Fig. 4  is for $M_{A}=2\cdot M_{Z}$ and $\tan\beta = 30$. As is seen, $|\sin\theta|$ hits unity before $m_{\tilde{t}_{1}}$
falls below $M_{Z}$. As in Fig. 2  $|\sin\theta|$ values as large as $|\sin\gamma|$ are reachable.

It may be convenient to have a comparative discussion of these four figures. That $m_{\tilde{t}_{1}}$ falls below $M_{Z}$
determines the starting of the forbidden regions in Figs. 1 and 3. In these figures $\tan\beta=2$ and $|\mu|$ term contribution 
is not suppressed at all, and $\Delta_{\tilde{t}}^{2}(\pi/2)$ grows rapidly with increasing $|A_{t}|$ and $|\mu|$ causing
$m_{\tilde{t}_{1}}$ to approach faster to $M_{Z}$. In these two figures $|\sin\theta|$ is controlled essentially by
$(|\mu||A_{t}|)/M_{A}^{2}$; however, $|\mu|$ and $|A_{t}|$ are constrained by charge and/or color breaking. Concerning this 
point one notes that the allowed range of $|\mu|$ and $|A_{t}|$ remain the same in both figures, and range of $|\sin\theta|$
follows form the value of $M_{A}$.

In Figs. 2 and 4 $\tan\beta=30$, and $m_{\tilde{t}_{1}}$ remains above $M_{Z}$ in a larger region than Figs. 1 and 3. This follows
simply from the $\cot\beta$ suppression of the $|\mu|$--term contribution to the stop mass--squared splitting
$\Delta_{\tilde{t}}^{2}(\pi/2)$. In both figures
starting of the forbidden region is determined by the fact that $|\sin\theta|$ exceeds unity. Hence, in both cases $|\sin\theta|$
assumes its allowed maximal value: $|\sin\theta|\sim |\sin\gamma|$. In sum, one observes that $\tan\beta$ plays a double role 
in determining $|\sin\theta|$: It both pushes away color and/or charge breaking boundary, and enhances $|\sin\theta|$ due to the 
dependence summarised in Item (3) in the previous page. 
 
The illustrations above show that there are certain regions of the MSSM parameter space in which $|\sin\theta|$ is
maximally large. As mentioned in the Introduction, this additional phase $\theta$ enters sfermion, chargino and neutralino 
mass matrices. To this end it is convenient to discuss first stop mass matrix. If one starts the whole analysis 
with a relative alignment between the Higgs doublets, as already mentioned in \cite{ben}, the entire effect is just a shift 
of $\gamma$ by $\theta$: $\gamma\rightarrow \gamma+\theta$ in the stop mass matrix. Thus, one may regard $\gamma$ at the 
right hand side of equation (3) as including $\theta$ already. In this sense reinsertion of $\theta$ to the stop squared-- 
mass matrix is inconsistent because $\theta$ itself is generated by $\gamma$ through (3). In sum, were it not for the
charginos and neutralinos one would just redefine the angle $\gamma$ with the replacement $\gamma\rightarrow
\gamma+\theta$, and a particular analysis of $\theta$ would be useless. In what follows presented is a detailed discussion 
of the CP violation in the chargino sector.
  
The charginos which are the mass eigenstates of charged gauginos and Higgsinos are described by a $2\times 2$ mass matrix 
\cite{kane}
\bea
M_{C}=\left(\begin{array}{c c} M_{2} & -\sqrt{2} M_{W} \cos \beta  \\
-\sqrt{2} M_{W} \sin \beta \; e^{i \theta} & |\mu| e^{i \varphi_{\mu}}\end{array}\right)
\eea
where $M_{2}$, which is the SU(2) gaugino mass, was made real already by appropriate field redefinitions mentioned 
in the Introduction \cite{phase1,phase2}. The masses of the charginos as well as their mixing matrices follow from the biunitary
transformation 
\bea
C_{R}^{\dagger} M_{C} C_{L} = \mbox{diag}(m_{\chi_{1}}, m_{\chi_{2}})
\eea
where $C_{L}$ and $C_{R}$ are $2\times 2$ unitary matrices, and $m_{\chi_{1}}$, $m_{\chi_{2}}$ are the masses of the 
charginos $\chi_{1}$, $\chi_{2}$ such that $m_{\chi_{1}}> m_{\chi_{2}}$. It is convenient to choose the following explicit
parametrization for the chargino mixing matrices:
\bea
C_{L}&=&\left(\begin{array}{c c} \cos \theta_{L} & \sin \theta_{L} e^{i \varphi_{L}}\\
- \sin \theta_{L} e^{- i \varphi_{L}} & \cos \theta_{L} \end{array}\right)\\
C_{R}&=&\left(\begin{array}{c c} \cos \theta_{R} & \sin \theta_{R} e^{i \varphi_{R}}\\
- \sin \theta_{R} e^{- i \varphi_{R}} & \cos \theta_{R} \end{array}\right) \cdot \left(\begin{array}{c c}
e^{i \phi_{1}} & 0 \\ 0 & e^{i \phi_{2}}\end{array}\right)
\eea
where the angle parameters $\theta_{L,R}$, $\varphi_{L,R}$, and $\phi_{1,2}$ can be determined from the defining equation (9).
A straightforward calculation yields 
\bea 
\tan 2 \theta_{L}&=& \frac{ \sqrt{8} M_{W} \sqrt{ M_{2}^{2} \cos^{2}\beta + |\mu|^{2} \sin^{2} \beta + |\mu| M_{2} \sin 2
\beta
\cos (\varphi_{\mu}-\theta)}}{M_{2}^{2}-|\mu|^{2}-2 M_{W}^{2} \cos 2\beta}\nonumber\\
\tan 2 \theta_{R}&=& \frac{ \sqrt{8} M_{W} \sqrt{ |\mu|^{2} \cos^{2}\beta + M_{2}^{2} \sin^{2} \beta + |\mu| M_{2} \sin 2
\beta  
\cos (\varphi_{\mu}-\theta)}}{M_{2}^{2}-|\mu|^{2}+2 M_{W}^{2} \cos 2\beta}\nonumber\\
\tan \varphi_{L}&=& \frac{ |\mu|\sin (\varphi_{\mu}-\theta)}{M_{2} \cot\beta + |\mu| \cos (\varphi_{\mu}-\theta)}\\
\tan \varphi_{R}&=& -\frac{ |\mu|\cot\beta \sin \varphi_{\mu}+ M_{2} \sin \theta}{|\mu| \cot \beta \cos \varphi_{\mu} + 
M_{2} \cos \theta}\nonumber
\eea
in terms of which the remaining two angles $\phi_{1}$ and $\phi_{2}$ read as follows
\begin{eqnarray}
\tan \phi_{i}= \frac{\mbox{Im}[Q_{i}]}{\mbox{Re}[Q_{i}]}
\eea
where $i=1,2$ and 
\bea
Q_{1}&=&\sqrt{2}M_{W}[\cos\beta \sin\theta_{L}\cos\theta_{R}e^{-i \varphi_{L}}+ \sin\beta \cos \theta_{L} \sin\theta_{R} e^{i
(\theta +\varphi_{R})}]\nonumber\\&+& M_{2} \cos\theta_{L}\cos\theta_{R} + |\mu| \sin\theta_{L} \sin\theta_{R} e^{i(
\varphi_{\mu}+\varphi_{R}-\varphi_{L})}\nonumber\\
Q_{2}&=&-\sqrt{2}M_{W}[\cos\beta \sin\theta_{R}\cos\theta_{L}e^{-i \varphi_{R}}+ \sin\beta \cos \theta_{R} \sin\theta_{L} e^{i
(\theta +\varphi_{L})}]\nonumber\\&+& M_{2} \sin\theta_{L}\sin\theta_{R}e^{i(\varphi_{L}-\varphi_{R})} + |\mu| \cos\theta_{L}
\cos\theta_{R} e^{i \varphi_{\mu}} \,.
\eea

The origin of the phases $\theta_{L,R}$, $\varphi_{L,R}$, and $\phi_{1,2}$ is easy to trace back. The angles $\theta_{L}$
and $\theta_{R}$ would be sufficent to diagonalize, respectively, the quadratic mass matrices $M_{C}^{\dagger} M_{C}$ and $M_{C}
M_{C}^{\dagger}$ if $M_{C}$ were real. As a result one needs the additional phases $\varphi_{L,R}$ which are identical to the
phases in the off--diagonal entries of the matrices  $M_{C}^{\dagger} M_{C}$ and $M_{C} M_{C}^{\dagger}$, respectively. However,
these four phases are not still sufficient for making the charginos masses real positive due to the biunitary nature of the
transformation (9); hence, the phases $\phi_{1}$ and $\phi_{2}$.

Inserting the unitary matrices $C_{L}$ and $C_{R}$ into the defining equation (9) one obtaines the following expressions
for the masses of the charginos
\bea
m^2_{\chi_{1(2)}}&=&\frac{1}{2}\Big\{M^2_2+|\mu|^2+2M^2_W +(-)
[(M^2_2-|\mu|^2)^2+4M^2_W\cos^{2}2\beta\nonumber\\
&+&4M^2_W(M^2_2+|\mu|^2+2M_2|\mu|\sin 2\beta \cos(\varphi_{\mu}-\theta))]^{1/2}\Big\}.
\eea  

As is clear from eqs. (8)-(14) all chargino mixing parameters as well as their masses themselves depend explicitly
on the phases $\varphi_{\mu}$ and $\theta$. From such dependencies of the derived quantities one immediately infers that:
\begin{itemize}
\item Even if $\varphi_{\mu}$ vanishes there is still a source for CP violation due to the presence of $\theta$.
\item If both $\varphi_{\mu}$ and $\varphi_{A_{t}}$ vanish then there remains no source for CP violation.
\item As the expressions of $\varphi_{R}$, $\phi_{1}$ and $\phi_{2}$ suggest, in general, it is not possible to absorb $\theta$
into a redefinition of $\varphi_{\mu}$. 
\end{itemize}

For an analysis of the contribution of $\theta$ to CP violation in chargino sector it is convenient to form appropriate
CP--violating quantities from the ones derived above. In general, when analyzing the effects of these phases on various observables, 
it is convenient assume a vanishing phase for the CKM matrix \cite{branco} to identify the pure supersymmetric contributions. 
In this context one recalls the discussions of $B\rightarrow X_{s} \gamma$ \cite{bsgam}, $B$--$\bar{B}$ and $K$--$\bar{K}$ mixings
\cite{bbar}, and EDM calculations (second and third references in \cite{side}) in which the CP violation due to the chargino 
contributions depends on the particular combinations of the matrices $C_{L}$ and $C_{R}$. Among all possible combinations, one can
choose, for example,  
\bea 
f_{i}(\theta,\varphi_{\mu})=\mbox{Im}\Big\{(C_{L}^{\dagger})_{i 2} (C_{R})_{1 i}\Big\}
\eea
where $i=1, 2$. In the actual applications the quantity in $\mbox{Im}\{...\}$ is multiplied by (1) the Yukawa 
couplings of the external fermions, (2) appropriate loop functions, and (3) the elements of the stop mixing matrix. 
The first two factors do not add new phases but weight the amount of CP violation. Especially the $\tan\beta$ dependence
of the Yukawa couplings is important. The third factor brings the additional phase $\varphi_{t}=\mbox{Arg}[A_{t}-\mu \cot\beta]$
which is important for determining the total CP--violation for a particular process. The net CP violation would be given by the first
two factors times the real part of the stop contribution times $f_{i}(\theta,\varphi_{\mu})$ plus the imaginary part of the stop
contribution times the real part of the chargino contribution. Since amount of the CP violating phase coming from the stop 
sector does not change with finite $\theta$, the main novelty (apart from $\theta$ dependence of various quantities such as the
stop and chargino masses) is brought about by $f_{i}(\theta,\varphi_{\mu})$. Keeping in mind these restrictions it is 
now time for having a numerical analysis of $f_{i}$ in the parameter spaces of Figs. 1-4.  

In the numerical study of $f_{i}(\theta,\varphi_{\mu})$ the free variable will be $\varphi_{\mu}$. The other parameters will 
be fixed from the parameter spaces of the Figs. 1 and 4. Thus it is convenient to introduce two parameter sets $A$ and $B$ 
\bea
A&=&\Big\{M_{\tilde{L}}=M_{\tilde{R}}=M_{2}= 2\cdot M_{Z},\; M_{A}=M_{Z},\; |A_{t}|=3.6\cdot M_{Z},\nonumber\\ &&|\mu|=6\cdot M_{Z},\; 
\tan\beta=2,\; \varphi_{A_{t}}=0.7\Big\}\,,\\ 
B&=&\Big\{M_{\tilde{L}}=M_{\tilde{R}}=M_{2}= 2\cdot M_{Z},\; M_{A}=2\cdot M_{Z},\; |A_{t}|=3\cdot M_{Z},\nonumber\\ &&|\mu|=3\cdot
M_{Z},\;
\tan\beta=30,\; \varphi_{A_{t}}=0.9 \Big\}\,,
\eea
subject to the constraints 
\bea
{\cal{C}}=\Big\{ m_{\tilde{t}_{1}}\geq M_{Z},\; m_{\chi_{2}}\geq M_{W},\; |\sin\theta|\leq 1\Big\}\,,
\eea 
for any $\varphi_{\mu}$ and $\varphi_{A_{t}}$.
Similar to the parameter spaces of Figs. 1-4 the mass parameters are chosen to lie slightly above $M_{Z}$. Such a choice for
the parameter space is motivated by the phenomenological requirements as well as minimization of the scale dependence of $\theta$. 
One notices that the values of the parameters $|A_{t}|$ and $|\mu|$ in $A$ and $B$ remain outside the allowed regions of Figs. 1 and
4, respectively. This does not pose a problem since these figures were formed for $\gamma=\pi/2$ whereas in the figures below CP  phases
will be varied over a range of values.  The specific choices for the phase $\varphi_{A_{t}}$ follow from the search strategy: One sets 
first $\varphi_{\mu}$ to zero and checks the variation of $f_{i}$ with $\varphi_{A_{t}}$ through its $\theta$ dependence. Generally, 
$f_{i}$ assumes a piece--wise continuous curve with $\varphi_{A_{t}}$ due to the violation of the consraints ${\cal{C}}$ place to
place. Then that value of $\varphi_{A_{t}}$ closest to the origin, and for which $f_{i}$ is maximum is picked up. In general,
there is no $a$ $priori$ condition saying what range of values are appropriate for $\varphi_{A_{t}}$. However, those values of the
supersymmetric phases as close as possible to a CP conserving point are interesting by themselves. In what follows, unless otherwise 
stated, all conditions in ${\cal{C}}$ will be applied irrespective of the presence or absence of the angle $\theta$ in the chargino
sector. That is, even if $\theta$ is not included in the chargino sector the condition on it in ${\cal{C}}$ will be applied in
illustrating $f_{i}$. 

Figs. 5-8 show the $\varphi_{\mu}$ dependence of $f_{i}$ for the parmeters in sets $A$ and $B$ in the presence (solid curves) and absence 
(dashed curves) of $\theta$ in the chargino sector. In each figure the dashed curve shows how big $f_{i}$ would be if there were no
contribution from the additional CP angle $\theta$. Moreover, in each figure argument of $f_{i}$ refers to the parameter sets $A$ or
$B$ as is self--explanatory. For the parameter set $A$, $f_{1}$ and $f_{2}$ are shown in Figs. 5 and 6, respectively. In each figure the
discontinuity at $\varphi_{\mu}\sim 0.3$ corresponds to the violation of any member of ${\cal{C}}$. 
There are such discontinuities at other values of $\varphi_{\mu}$ as well but for clarity their discussion will be given later.
As expected, in the absence of $\theta$, in each figure $f_{i}$ increases linearly with $\varphi_{\mu}$ 
until ${\cal{C}}$ is violated. However, due to $\varphi_{A_{t}}$ support at finite $\theta$, $|f_{1}|$  starting from its maximum at
$\varphi_{\mu}=0$ decreases gradually until $\varphi_{\mu}\sim 0.3$. Except for a small region between $\varphi_{\mu}=0.27$ and
$\varphi_{\mu}=0.3$ $|f_{1}|$ for finite $\theta$ is larger than that for vanishing $\theta$. By construction, the figure suggests that
$|f_{1}|$ is quite large for vanishing $\varphi_{\mu}$. Fig. 6 shows the behaviour of $f_{2}$ as in Fig. 5. In this case, solid 
curve is much larger than the dashed one everywhere, in particular, the solid curve starts $|f_{2}|\sim 0.17$ at $\varphi_{\mu}=0$ and
increases near $0.2$ at $\varphi_{\mu}\sim 0.2$ beyond which it decreases gradually. Taking the arithmetic means of the curves in the
relevant $\varphi_{\mu}$ interval one observes that there is an order of magnitude enhancement. The interesting thing is that for
larger values of $\varphi_{\mu}$ the two curves become at most equal implying that the finite $\theta$ contribution is important. 

Similar to Figs. 5 and 6, depicted in Figs. 7 and 8 are the $\varphi_{\mu}$--dependence of $f_{1}$ and $f_{2}$ for the parameter set
$B$. Here it is seen that the threshold value of $\varphi_{\mu}$ at which ${\cal{C}}$ is violated is approximately one order of
magnitude below the one for the parameter set $A$. Hence in the former the $\varphi_{\mu}$ values are closer to the CP conserving
points $\varphi_{\mu}=0$ than the latter. This stems mainly from the slightly large value of $\tan\beta$ which pushes 
$|\sin\theta|$ beyond its limits in ${\cal{C}}$. This effect could also be observed in Figs. 2 and 4. Again with the mean values of the 
curves one obtains more than an order of magnitude enhancement in $f_{i}$ for finite $\theta$. One notices that, in all four figures,
there is no sign shift in $f_{i}$ for finite and vanishing $\theta$. However, relative magnitudes of $f_{i}$ change in finite  $\theta$
case. 
 
Altough $\varphi_{\mu}$ values involved in Figs. 5-8 are relatively small, the $\varphi_{A_{t}}$ value is close to unity, and thus, it is 
quite away from the CP--conserving point $\varphi_{A_{t}}=0$. Moreover, one may wonder if $f_{i}$ in the presence of $\theta$ does really
dominate over the one in the absence of $\theta$ in the chargino sector. This point is important because it may happen that $f_{i}$ for
finite $\theta$ could  be of similar magnitude as the one for vanishing $\theta$ so that the net effect of introducing $\theta$ could be
just a shift in the value of $\varphi_{\mu}$. Altough the expressions for various angles in eqs. (12)-(14) guarantee that this does not
happen, it may still be convenient to illustrate this point for a sample parameter space. In doing this it is appropriate 
to search for a region of the parameter space where both CP phases remain in close vicinity of a CP--conserving point. In this
respect one can introduce the following parameter set:

\bea
C&=&\Big\{M_{\tilde{L}}=M_{\tilde{R}}=M_{2}= 4\cdot M_{Z},\; M_{A}=M_{Z},\; |A_{t}|=11\cdot M_{Z},\nonumber\\ &&|\mu|=11\cdot M_{Z},\;
\tan\beta=2,\; \varphi_{A_{t}}=0.052\Big\}
\eea
which is again subject to the constraint ${\cal{C}}$. It is obvious that now stop left--right mixings are chosen to be large 
in accordance with stop soft mass parameters: $|A_{t}|=|\mu|\sim \mbox{a}\, \mbox{TeV}$, and $M_{\tilde{L}}=M_{\tilde{R}}\sim
350\,\mbox{GeV}$. Before discussing the implications of the set $C$, it is convenient to see, for example,  $f_{1}$ in Fig. 7 
in the entire range of $\varphi_{\mu}$ to observe the competetion between values of $f_{1}$ when $\theta$ is present and absent. Fig. 9
shows the dependence of $f_{1}$ on $\varphi_{\mu}$ for the parameter set $B$ for finite (solid curve) and vanishing (dashed curve)
$\theta$ in the chargino mass matrix. First one observes the absence of any symmetry in the behaviour of $f_{1}$ which is already
expected from the interference among various sinus and cosinus functions. The next thing is that away from the CP--conserving points,
$[\varphi_{\mu}=0,\pi,2\pi]$, $f_{1}$
for vanishing $\theta$ dominates and assumes approximately the same value that it gets for finite $\theta$ near the CP--conserving
points. Assuming that there is no constraint on the range of $\varphi_{\mu}$, one observes that the net effect of $\theta$ is to
push those large CP violation points near CP conserving ones. For instance, $f_{1}[\varphi_{\mu}\sim 1.5, \theta=0]\equiv
f_{1}[\varphi_{\mu}\sim \pi, \theta\neq 0]$, so that it is just a matter of shift in the value of $\varphi_{\mu}$. However, there
are counter examples to this conclusion where $f_{1}$ for finite $\theta$ dominates on the one for vanishing $\theta$. The first 
example is based on the parameter set $C$ above whereas the other two are dicussed below for the parameter sets $D$ and $E$.
The parameter set $C$ suggests a rather small value for the threshold value of $\varphi_{A_{t}}$ compared to the previous sets $B$ and
$A$. Hence one is comparatively closer to the CP conserving point $\varphi_{A_{t}}=0$. In this context, Fig. 10 shows the
dependence of $f_{1}$ on $\varphi_{\mu}$ for the parameter set $C$ for finite (solid curve) and vanishing (dashed curve) $\theta$.
It is obvious that nowhere dashed curve appears, and this stems from the fact that $f_{1}$ for vanishing $\theta$ is small.
To see this better, one may check Fig. 11 where variation of $f_{1}$  near  $\varphi_{\mu}=2\pi$ is shown for finite (solid curve) and
vanishing
(dashed curve) $\theta$. As the figure suggests $f_{1}$ at finite $\theta$ is $\sim 6$ times larger than the one for vanishing
$\theta$ when the latter attains its maximum. However, one notices that the parameter set $C$ already pushes the non-vanishing $f_{1}$
region near a CP--conserving point around which such a dominance is already expected. 

In Figs. 5-11 the main interest has been the $\varphi_{\mu}$--dependence of $f_{i}$ for fixed values of the remaining parameters as
given in the sets $A$, $B$ and $C$. The main conclusion from these examples is the $existence$ $of$ $large$ $\mbox{CP}$ $violation$
$near$ $the$ $\mbox{CP}$--$conserving$ $points$ where $f_{i}$, with $\theta$ present, can, depending on the parameter space,  dominate
on the one when $\theta$ is absent in certain portions of or everywhere along the $\varphi_{\mu}$ axis. 

The next example to show the dominance of $f_{i}$ at finite $\theta$ over the ones at vanishing $\theta$ consists of the variation of
$f_{i}$ with $\tan\beta$ and $M_{A}$. For this purpose, it is convenient to introduce two new parameter sets
 
\bea
D&=&\Big\{M_{\tilde{L}}=M_{\tilde{R}}=M_{2}= 2\cdot M_{Z},\; M_{A}=M_{Z},\; |A_{t}|=1.6\cdot M_{Z},\nonumber\\ &&|\mu|=1.6\cdot M_{Z},\;
\varphi_{\mu}=0.5,\; \varphi_{A_{t}}=0.5\Big\}\,,\\
E&=&\Big\{M_{\tilde{L}}=M_{\tilde{R}}=M_{2}= 2\cdot M_{Z},\; M_{A}=4\cdot M_{Z},\; |A_{t}|=1.6\cdot M_{Z},\nonumber\\ &&|\mu|=1.6\cdot
M_{Z},\;
\varphi_{\mu}=0.5,\; \varphi_{A_{t}}=0.5\Big\}\,.
\eea
Contrary to Figs. 5-11 in the following figures the constraint $|\sin\theta|\geq 1$ will  be relaxed when there is no $\theta$
contribution to the chargino mass matrix; namely, the constraints on $\theta$ will be just forgotten when analyzing the bare chargino
mass matrix. This is motivated by the requirement of observing the relative magnitudes of $f_{i}$ as a function of $\tan\beta$
when $\theta$ is present and absent. Moreover, in the following figures $\tan\beta$ will be varied from 2 to 100 mainly for observing 
the asymptotic behaviour of the chargino mixings when there is no $\theta$ contribution. However, the derivation $\theta$ in eq. (3)
\cite{ben} assumes the dominance of the top quark and top squark loops, so that graphs of $f_{i}$ when $\theta$ is present sould be taken
with great $care$ for $\tan\beta\simgt 40$. In this respect, it is healty to compare the asymptotic value of $f_{i}$ for vanishing
$\theta$ with $f_{i}$ for finite $\theta$ below $\tan\beta\sim 40$.  

Depicted in Fig. 12 is $\tan\beta$--dependence of $f_{1}$ for the parameter set $D$ when $\theta$ is present (solid curve) and absent
(dashed curve) in the chargino
mass matrix. For small $\tan\beta$ the two curves are close to each other due to the fact that $|\sin\theta|$ is small. But then, as
$\tan\beta$ increases, the solid curve rapidly increases, and becomes twice bigger than the dashed curve. Beyond $\tan\beta\sim
20$ the solid
curve vanishes because of the violation of $|\sin\theta|$ constraint in ${\cal{C}}$. However, as mentioned in the previous paragraph the
dashed curve keeps increasing slowly such that even at $\tan\beta=100$ it remains around its value at $\tan\beta=2$. In particular, it can
never chatch the solid curve with such a slow increase. It is seen that $|\sin\theta|$ forces the solid curve to vanish before the
validity limit of eq. (3) is violated. Essentially the same thing happens for $f_{2}$ in Fig. 13. Additionally, to see the effects of
$M_{A}$ on $f_{i}$, one can use the set $E$ in which case $\sin\theta$ in eq. (3) is reduced by a factor of 16 compared to the one in set
$D$. Needless to
say, all dased curves will keep their behaviour in Figs. 12 and 13. In this respect, one can see Figs. 14 and 15 where, in accordance
with large $M_{A}$ value in the set $E$, the dashed and solid curves get closer at $\tan\beta=2$. Moreover, the solid curves nowhere
vanish because the constraint ${\cal{C}}$ remains unviolated due to large $M_{A}$.  Even at $\tan\beta=100$ for dahsed curve it is 
impossible to chatch any value of the solid curve below $\tan\beta\sim 40$. Hence the conclusion remains the same as for Figs. 12 and 13. 
As a result, the study of Figs. 12-15 adds one more conclusion: $value$ $of$ $f_{i}$, $when$ $\theta$ $is$ $present$, $remains$ $much$
$larger$ $than$ $the$ $value$ $of$ $f_{i}$ $when$ $\theta$ $is$ $absent$, $and$ $this$ $gap$ $cannot$ $be$ $closed$ $however$ $large$
$\tan\beta$ $is$ $chosen$.

The analytic as well as the numerical analyses above have been based solely on the chargino sector for simplicity. Of course, 
the neutralino sector is also important and must be analyzed in detail. The method of discussion in that case proceeds on similar lines
with mostly numerical procedures because the neutralinos, which are the mass eigenstates of neutral gauginos and Higgsinos, are
described by a $4\times 4$ mass matrix whose analytic treatment is not illuminating. However, just because of the similarity of
the mass matrices in terms of their dependencies on $\mu$ parameter and $\theta$ \cite{spon} one expects similar effects to take
place in the case of neutralinos, too. Though not discussed here,  effects of the additional phase $\theta$ on the neutralino
sector are no way negligible, as they are equially important for the phenomena where charginos play a role. After tracing the 
effects of $\theta$ on the chargino sector one infers the neutralino sector be affected by similar ways, and their treatment 
can be done in discussing the specific processes where neutralinos contribute.

When discussing the Figs. 5-11 importance of proximity to a CP conserving point has been frequently emphasized without an explicit 
reason. Therefore it may be convenient to have a detailed discussion of it. First, one recalls the situation in minimal supergravity models
with finite phases. The phase of the $\mu$ parameter remains unchanged, and the phase of the Higgs--sfermion trilinear 
couplings gets aligned towards that of the gaugino masses during the RGE running from GUT scale to the weak scale. Thus,
the $A$--terms cannot have a large phase after field redefinitions to obtain the physical set of phases, and this remains true
unless one chooses the universal gaugino mass much smaller than the universal $A$--terms at the supergravity scale. On
the other hand particle EDM's constrain the $\mu$ parameter severely. In this sense, 
in mimimal supergravity models (see the second reference in \cite{others}) the low energy model is necessarily close to
a CP--conserving point. The second reason for being in the vicinity of the CP--conserving points follows from the 
relaxation property of the pseudo--Goldstone modes of the U(1)$_{PQ}$ and U(1)$_{R-PQ}$ symmetries of the MSSM Lagrangian 
in a generalization of the Peccei--Quinn mechanism. It has been shown in \cite{phase2} that these Goldstone modes
relax to or near a CP--conserving point leading to a solution of the strong and supersymmetric CP problems 
by similar mechanisms. Furthermore, specific realizations of these Goldstone modes have already been constructed in dynamical 
supersymmetry breaking scenarios \cite{sugra}. One of the major effects of these phases would be the induction of 
a macroscopic Yukawa-type force between the massive bodies \cite{phase2} which can be significant for phases ${\cal{O}}(1)$
since the masses of the pseudo--Goldstone modes can lie just above $a$ $\mbox{TeV}$. In this sense, one expects MSSM be near a
CP--conserving  point after the general derivation in \cite{phase2}, and thus, $having$ $large$ CP $violation$ $when$ $\varphi_{A_{t}}$
$and$ $\varphi_{\mu}$ $are$ $near$ $a$ CP--$conserving$ $point$ $is$ $important$. 
  
Even if one assumes a typical globally supersymmetric low energy model with the SM gauge group by just forgetting about the 
constraints in the previous paragraph, still there are nontrivial effects of the additional phase $\theta$ on the phenomena 
where $\varphi_{A_{t}}$ and $\varphi_{\mu}$ play a role. For example, one recalls the cancellation of different supersymmetric
contributions in certain portions of the MSSM parameter space (See the second and third references in \cite{side}) in calculating
the EDM's or other FCNC transitions. As the graphs in Figs. 12-15 show clearly, now the $\tan\beta$--sensitivity of the supersymmetric
phases is modified; hence, the constraints on the MSSM parameter space. Another important point deserving discussion concerns the
electroweak baryogenesis \cite{baryo} where the necessary amount of CP--violation could be of explicit or spontaneous \cite{spon}
in nature though the latter can require an unrealistically small pseudoscalar mass. The observation here is that the relative alignment
between the Higgs doublets, even for very small $\varphi_{\mu}$ and $\varphi_{A_{t}}$, can be large enough to source necessary CP
violation for the electroweak baryogenesis.
 
\section{Conclusion}
This work has dealth with the induction and subsequent effects of the unremovable alignment between the Higgs doublets due to explicit
CP violation in the MSSM Lagrangian. This radiatively induced phase could be as large as the CP phases themselves in easy--to--find
portions of the MSSM parameter space as examplified by Figs. 1-4. This additional phase can affect chargino and neutralino sectors 
through their mass matrices. In here, for the particular case of charginos, the effects of this phase have been discussed both
analytically 
and numerically. It is shown that this additional phase can induce large CP violation near the CP--conserving points in the chargino
sector. Moreover, the CP violation in the presence of this radiatively induced phase is highly sensitive to $\tan\beta$, and larger
than the one in the absence of it. Altough no discussion of the neutralinos is given, just by the similarity of the mass matrices in
terms
of their dependence on this extra phase  one expects neutralino sector to have similar additional CP violation effects .

Now it is of phenomenological interest to determine the additional CP violation effects on various phenomena where the supersymmetric 
CP violation effects contribute. For example, an analysis of the neutralino sector together with the results of this work will be
important for  EDM's of particles \cite{edm}, LSP searches \cite{falk},
$K$-- and $B$-- physics \cite{phase1,others}, electroweak baryogenesis \cite{baryo}, and  weak and
electromagnetic dipole formfactors \cite{dipole}. In particular, for a real CKM matrix \cite{branco}, the CP--violation in 
$K$ and $B$ physics \cite{phase1,others} follows solely from the chargino and neutralino sectors so that this additional 
phase $\theta$ becomes important there. 

\section{Acknowledgements}
The author is grateful to Antonio Masiero, Mariano Quir{\'o}s, and Oscar Vives for fruitful discussions. 
  
\newpage
\begin{figure}
\setlength{\epsfxsize}{10cm}
\centerline{\epsfbox{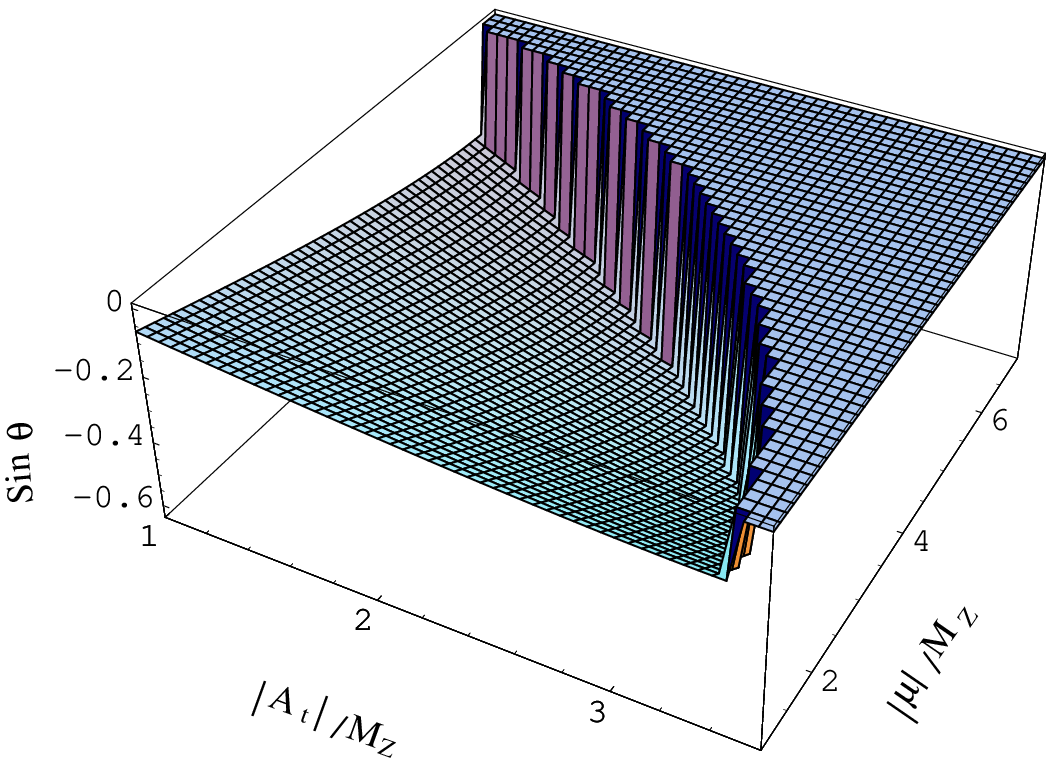}}
\caption{$\sin\theta$ in $|A_{t}|/M_{Z}$ -- $|\mu|/M_{Z}$ plane for $\tan\beta=2$ and $M_{A}=M_{Z}$.
The $\sin\theta=0$ plateau $is$ $not$ the value of $\sin\theta$; it shows just the parameter region
$excluded$ by either $m_{\tilde{t}_{1}}< M_{Z}$ or $|\sin\theta|> 1$.}
\end{figure}

\begin{figure}
\setlength{\epsfxsize}{10cm}
\centerline{\epsfbox{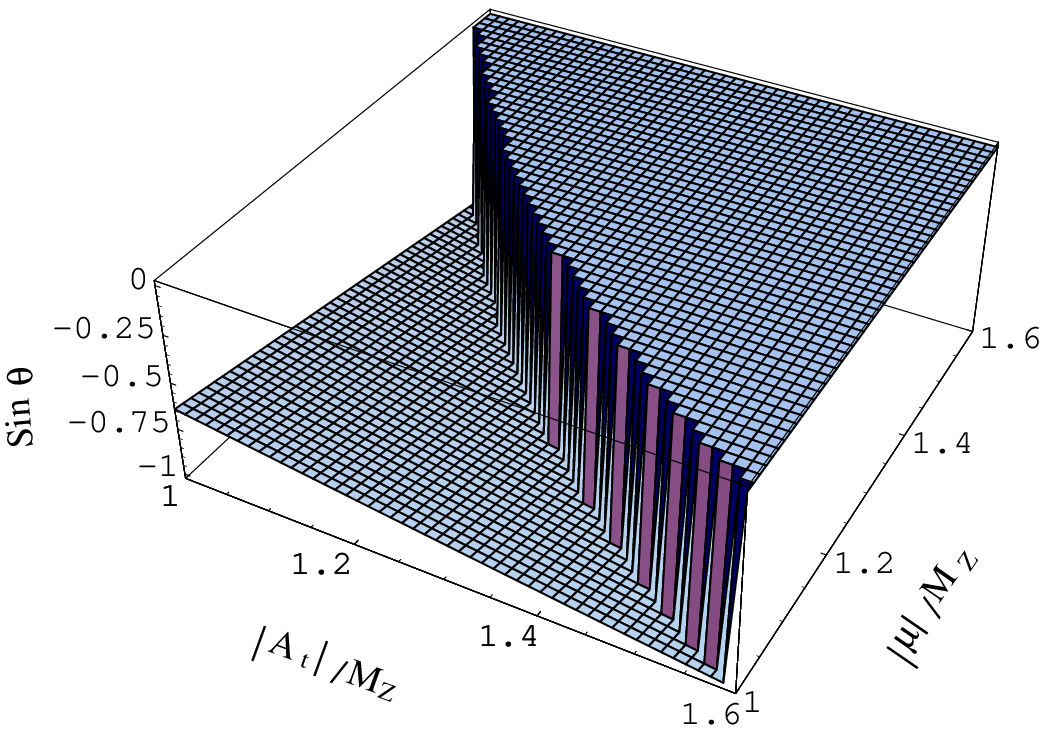}}
\caption{The same as in Fig. 1, but for $M_{A}=M_{Z}$ and $\tan\beta=30$.}
\end{figure}

\begin{figure}
\setlength{\epsfxsize}{10cm}
\centerline{\epsfbox{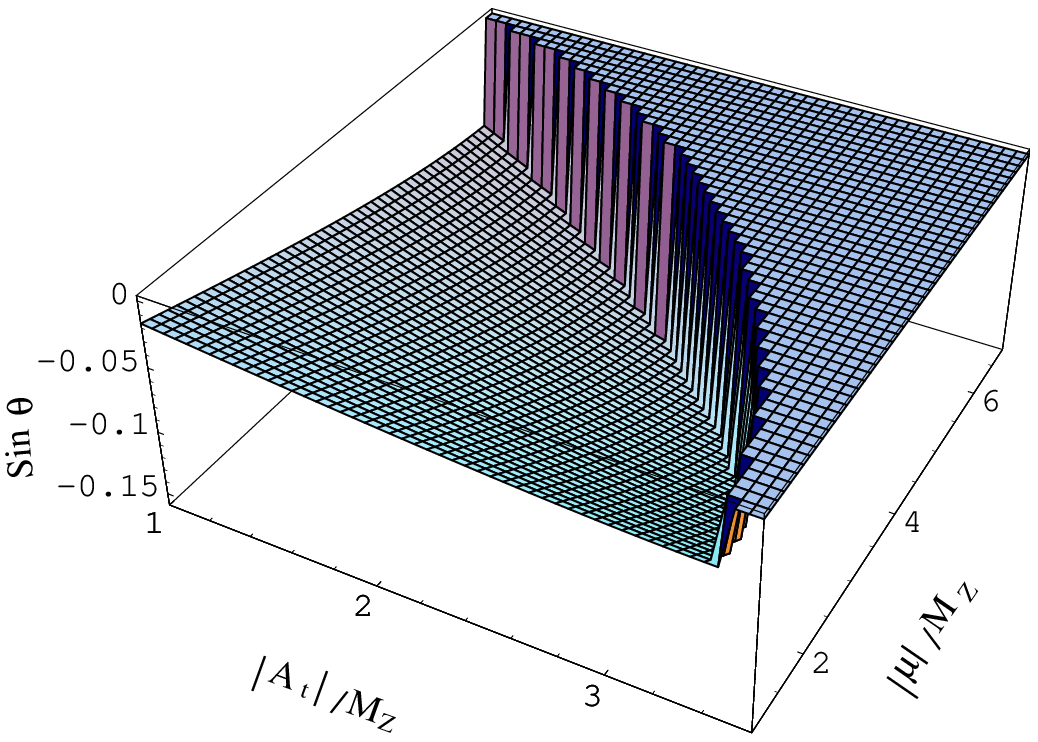}}
\caption{The same as in Fig. 1, but for $M_{A}=2\cdot M_{Z}$ and $\tan\beta=2$.}
\end{figure}

\begin{figure}
\setlength{\epsfxsize}{10cm}
\centerline{\epsfbox{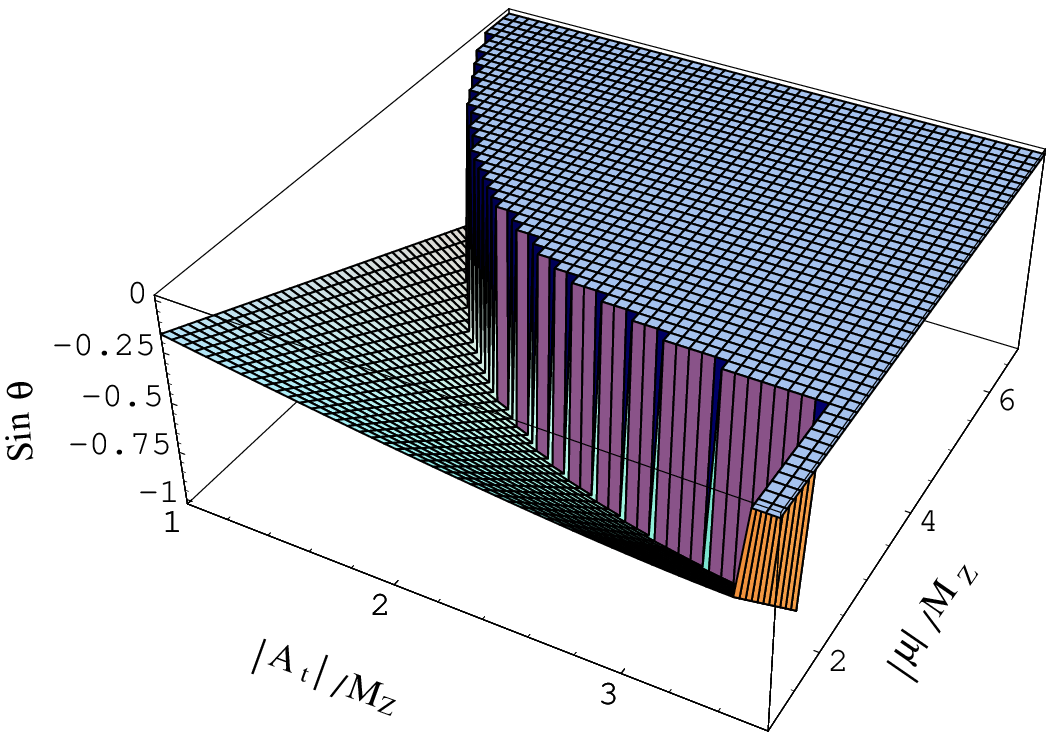}}
\caption{The same as in Fig. 1, but for $M_{A}=2\cdot M_{Z}$ and $\tan\beta=30$.}
\end{figure}

\begin{figure}
\setlength{\epsfxsize}{10cm}   
\centerline{\epsfbox{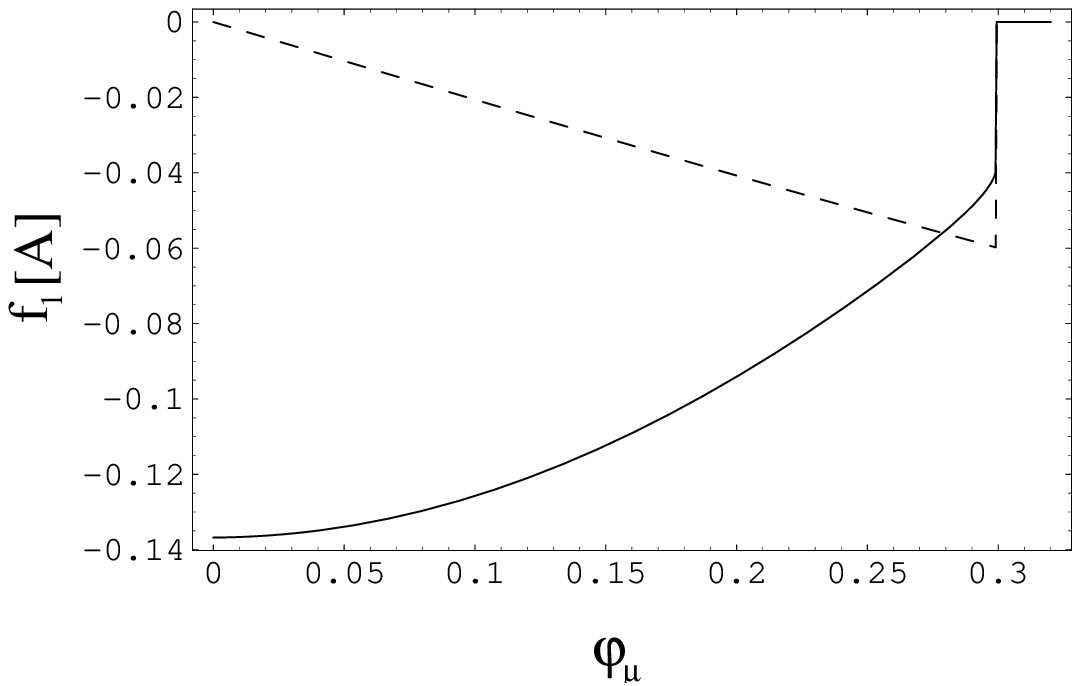}}
\caption{The dependence of $f_{1}$ on $\varphi_{\mu}$ for the parameter set $A$.}
\end{figure}

\begin{figure}
\setlength{\epsfxsize}{10cm}
\centerline{\epsfbox{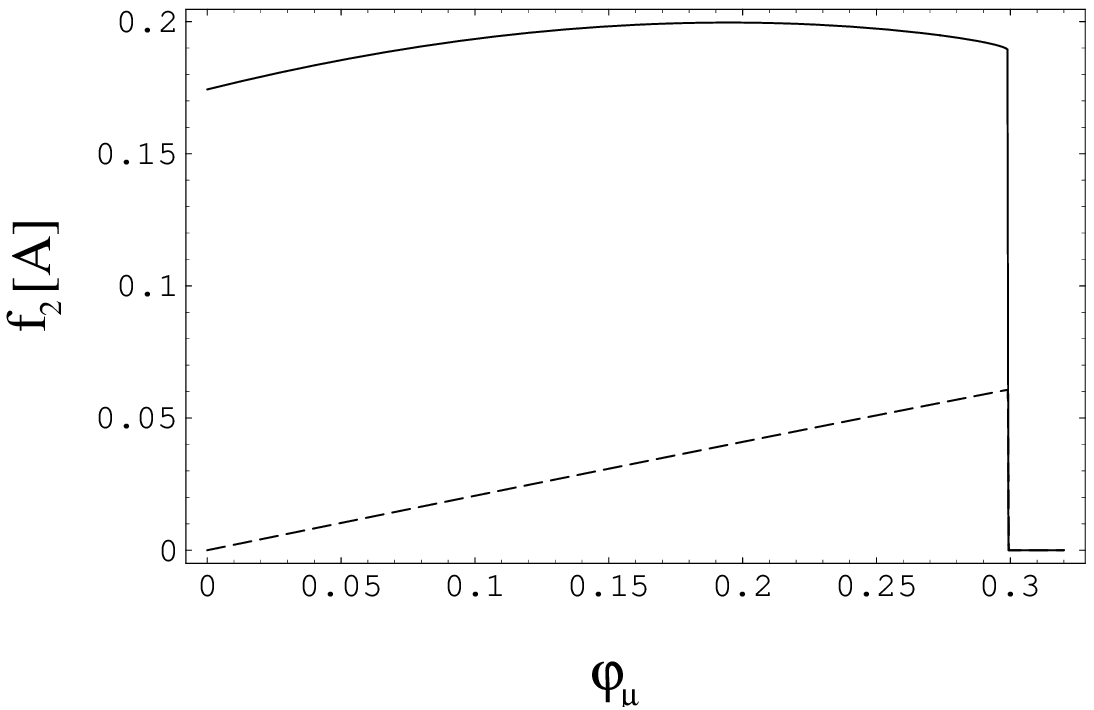}}
\caption{The dependence of $f_{2}$ on $\varphi_{\mu}$ for the parameter set $A$.}
\end{figure}

\begin{figure}
\setlength{\epsfxsize}{10cm}
\centerline{\epsfbox{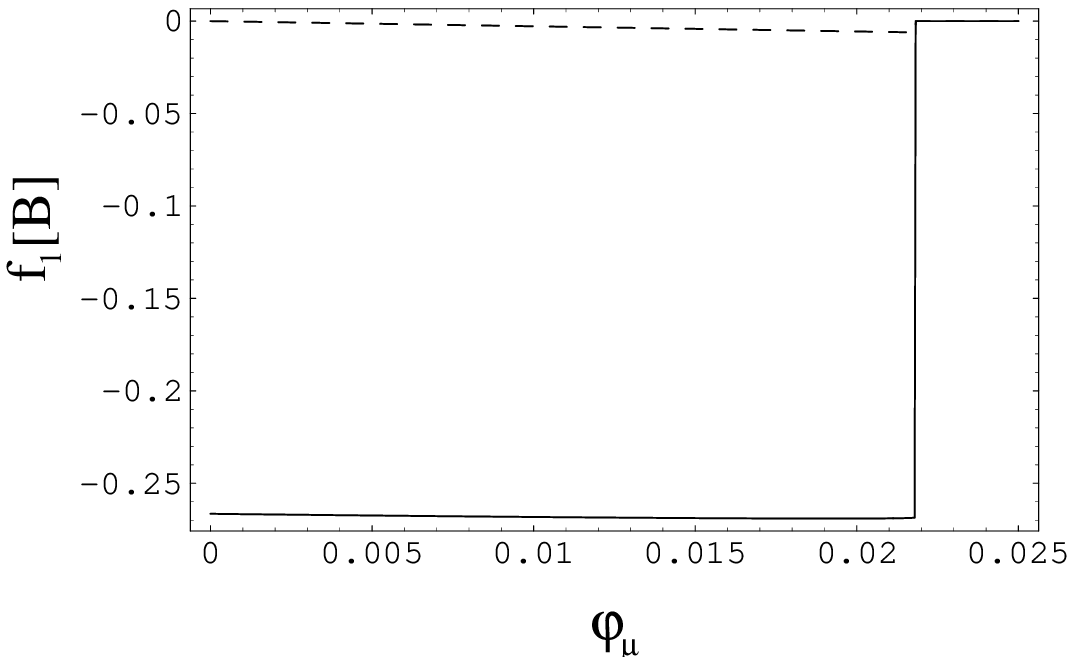}}
\caption{The dependence of $f_{1}$ on $\varphi_{\mu}$ for the parameter set $B$.}
\end{figure}

\begin{figure}
\setlength{\epsfxsize}{10cm}
\centerline{\epsfbox{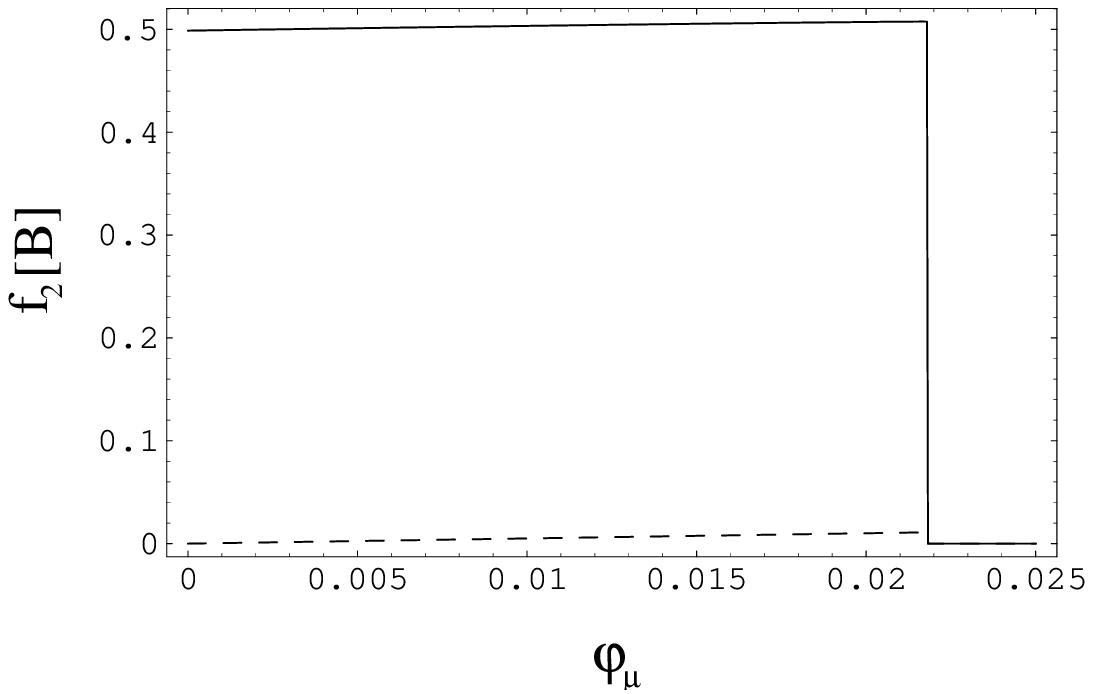}}
\caption{The dependence of $f_{2}$ on $\varphi_{\mu}$ for the parameter set $B$.} 
\end{figure}

\begin{figure}
\setlength{\epsfxsize}{10cm}   
\centerline{\epsfbox{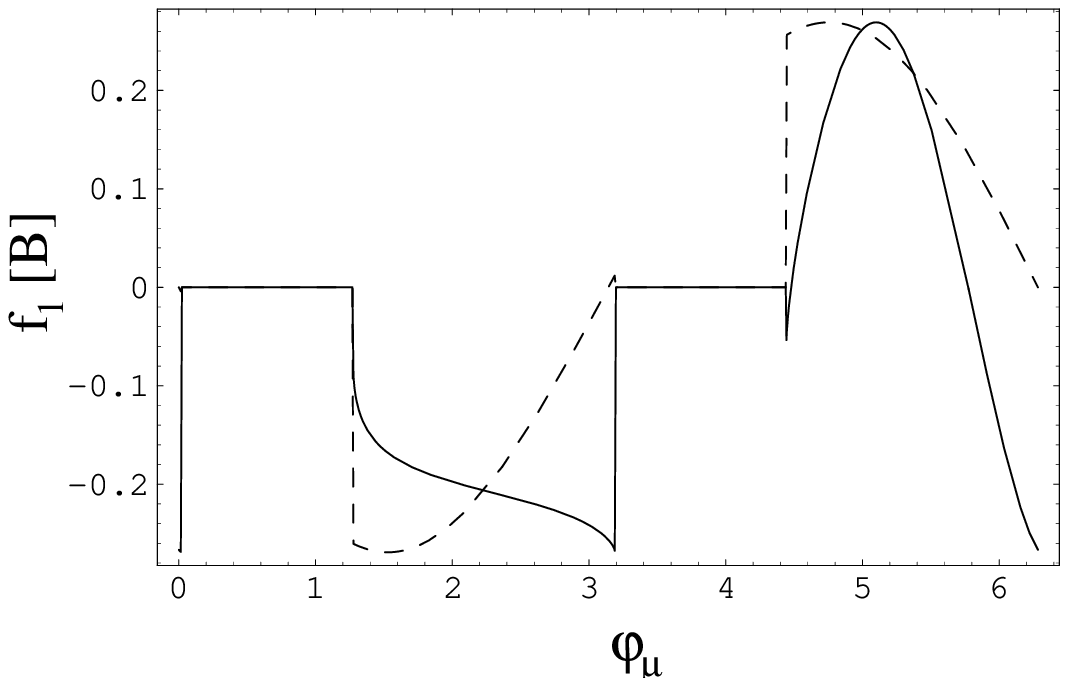}}
\caption{$f_{1}$ for the parameter set $B$ in the entire range of $\varphi_{\mu}$.} 
\end{figure}

\begin{figure}
\setlength{\epsfxsize}{10cm}
\centerline{\epsfbox{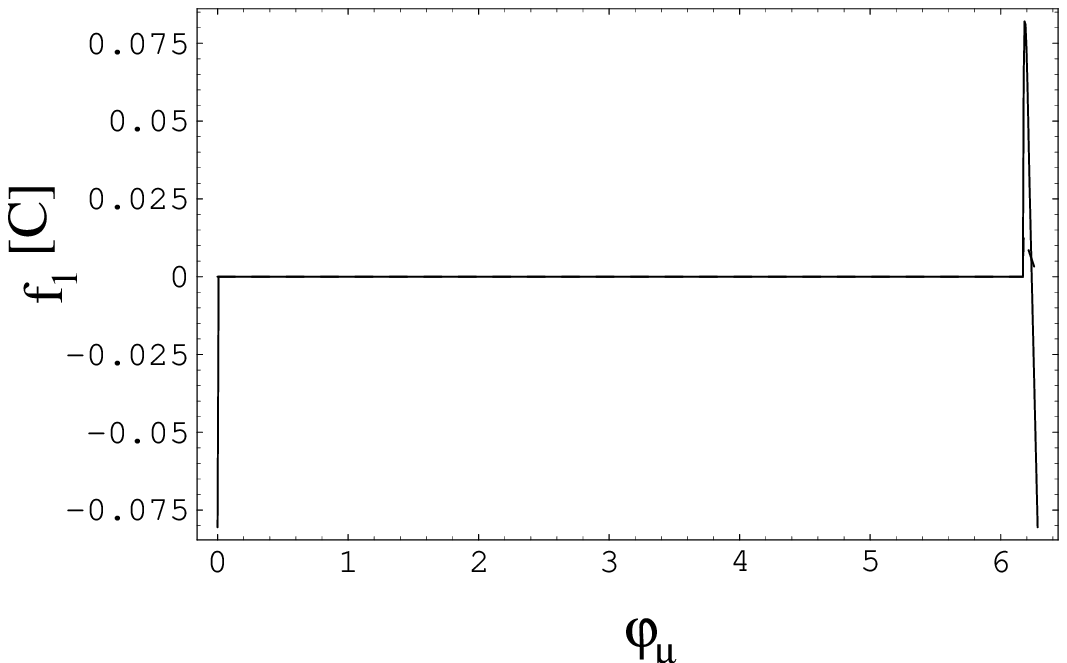}}
\caption{$f_{1}$ for the parameter set $C$ in the entire range of $\varphi_{\mu}$.}
\end{figure}

\begin{figure}
\setlength{\epsfxsize}{10cm}
\centerline{\epsfbox{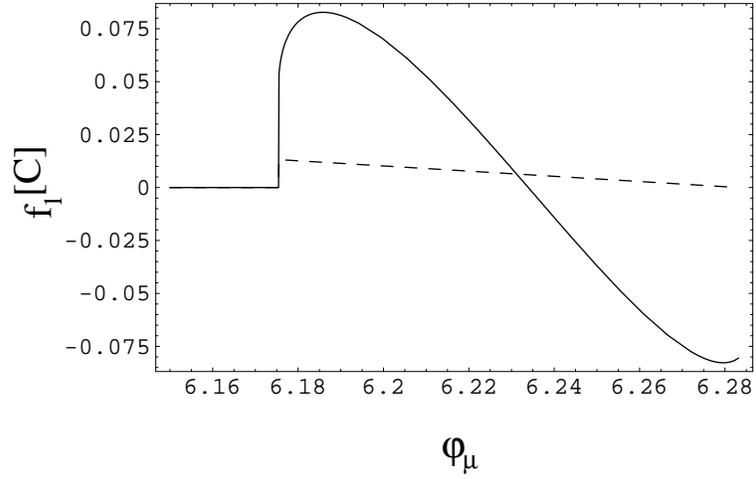}}
\caption{$f_{1}$ for the parameter set $C$ in the vicinity of $\varphi_{\mu}=2\pi$.}
\end{figure}

\begin{figure}
\setlength{\epsfxsize}{10cm}
\centerline{\epsfbox{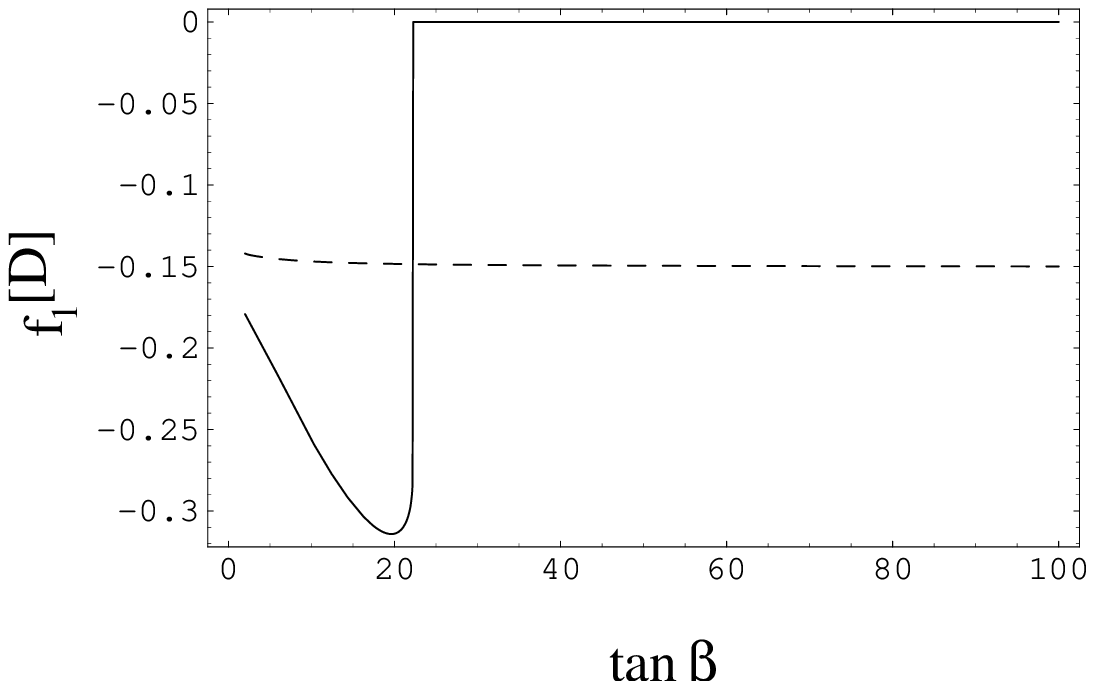}}
\caption{Variation of $f_{1}$ with $\tan\beta$ for the parameter set $D$.}
\end{figure}

\begin{figure}
\setlength{\epsfxsize}{10cm}
\centerline{\epsfbox{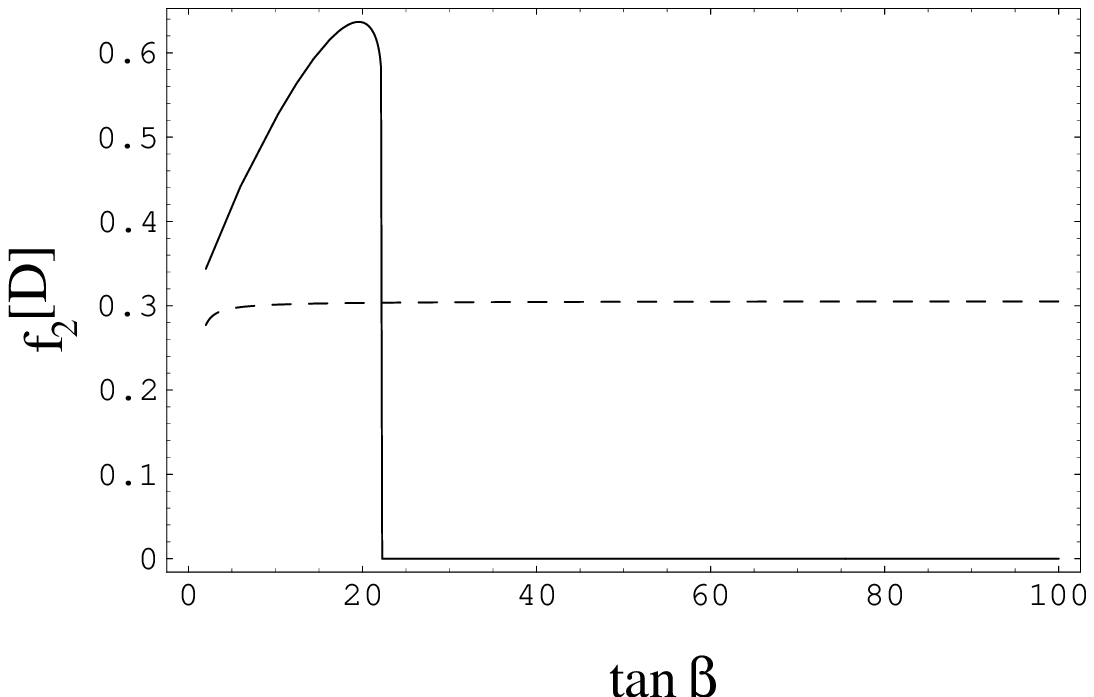}}
\caption{Variation of $f_{2}$ with $\tan\beta$ for the parameter set $D$.}
\end{figure}

\begin{figure}
\setlength{\epsfxsize}{10cm}
\centerline{\epsfbox{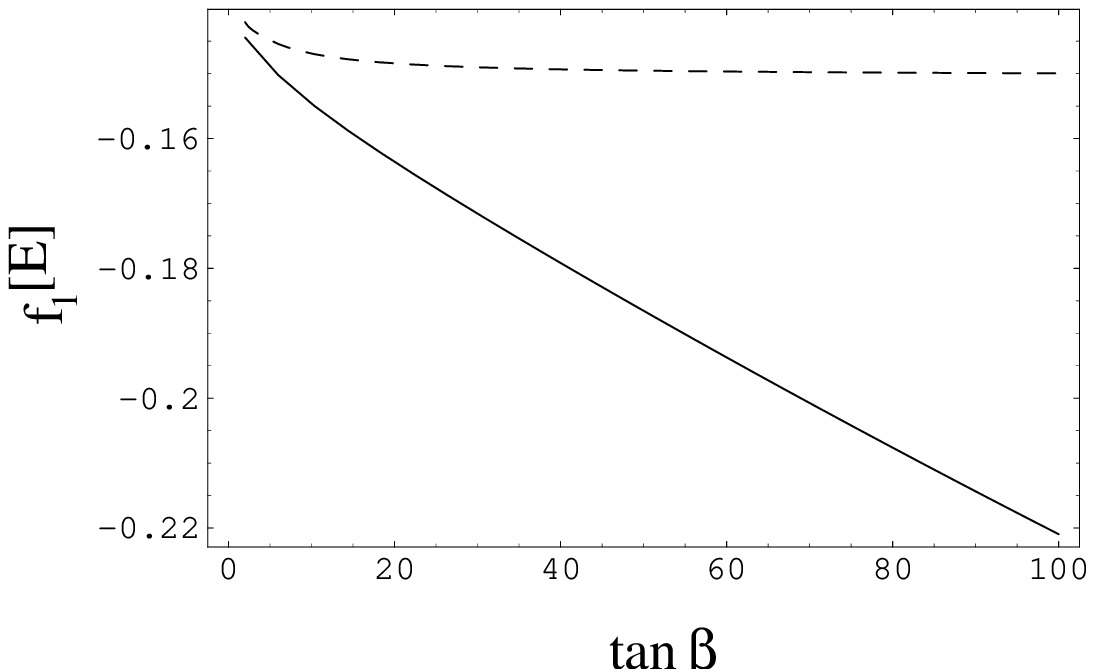}}
\caption{Variation of $f_{1}$ with $\tan\beta$ for the parameter set $E$.}
\end{figure}

\begin{figure}
\setlength{\epsfxsize}{10cm}
\centerline{\epsfbox{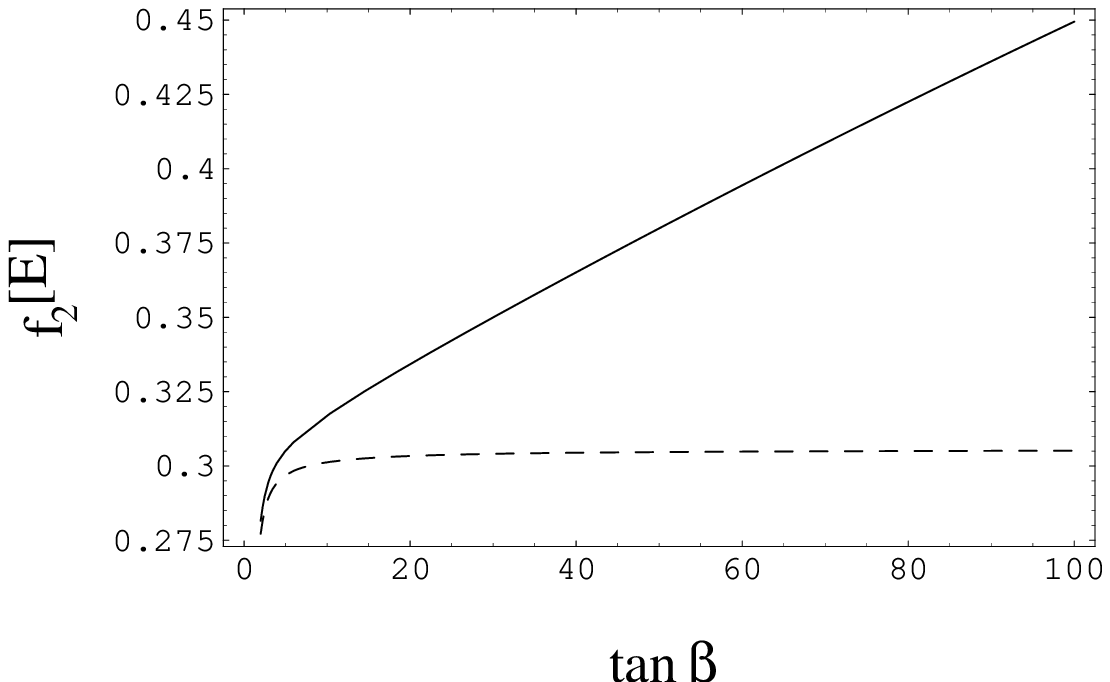}}
\caption{Variation of $f_{2}$ with $\tan\beta$ for the parameter set $E$.}
\end{figure}

\end{document}